# Deep MRI Reconstruction: Unrolled Optimization Algorithms Meet Neural Networks

Dong Liang#*, *Senior Member, IEEE*, Jing Cheng#, *Student Member, IEEE,* Ziwen, Ke, Leslie Ying*, *Senior Member, IEEE*

*Abstract*—Image reconstruction from undersampled k-space data has been playing an important role for fast MRI. Recently, deep learning has demonstrated tremendous success in various fields and has also shown potential to significantly speed up MRI reconstruction with reduced measurements. This article gives an overview of deep learning-based image reconstruction methods for MRI. Three types of deep learning-based approaches are reviewed, the data-driven, model-driven and integrated approaches. The main structure of each network in three approaches is explained and the analysis of common parts of reviewed networks and differences in-between are highlighted. Based on the review, a number of signal processing issues are discussed for maximizing the potential of deep reconstruction for fast MRI. The discussion may facilitate further development of "optimal" network and performance analysis from a theoretical point of view.

## I. INTRODUCTION

Since its inception in the early 70's, magnetic resonance imaging (MRI) has revolutionized radiology and medicine. However, MRI is known to be a slow imaging modality and many techniques have been developed to reconstruct the desired image from undersampled measured data to improve the imaging speed [1]. During the past decades, compressed sensing (CS) has become an important strategy for fast MR imaging based on the sparsity prior. However, the iterative solution procedure takes a relatively long time to achieve a high-quality reconstruction, and the selection of the regularization parameter is empirical. Although some numerical methods such as Stein's Unbiased Risk Estimation (SURE) [2] have been proposed to optimize the free parameters in MR imaging, these methods are burdened with high computational complexity. Additionally, most approaches only exploit prior information either directly from the to-be-reconstructed images or with very few reference images involved.

Recently, deep learning has demonstrated tremendous success and become a growing trend in general data analysis [3-5]. Inspired by such success, deep learning has been applied to computational MRI and shown potential to significantly speed up MR reconstruction [6-26]. In contrast to compressed sensing and other constrained reconstruction methods, deep learning avoids complicated optimization-parameter tuning and performs superfast online reconstruction, with the aid of offline training using enormous data.

Deep learning–based MRI reconstruction methods can be roughly divided into three categories: data-driven approaches [6-16], model-driven approaches [23-26] and integrated approaches [17-22]. Data-driven approaches use the standard network architecture as a "black box" to learn the mapping between the input and output data, heavily relying on a huge amount of data to train the black box without utilization of any prior knowledge. Advanced network architectures have been investigated to increase the learning capability of data-driven methods. Model-driven approaches are based on the constraint reconstruction model and unroll the procedure of an iterative optimization algorithm to a deep network, while learning the parameters and functions through network training. Model-driven deep learning methods have the advantage of model-based CS approaches that the solution space is reduced by the model, but also avoid the limitation of model-based approaches that the model might be oversimplified. As a result, such networks usually perform well with a smaller size of training sets. Table 1 summarizes the difference and relationship between these two approaches. Methods that include the ingredients of "unrolling" and "black box" from both approaches are classified as integrated approaches.

The main purpose of this article is to give an overview of deep learning-based MR image reconstruction methods, with an effort to highlight their unique properties and connections between them. We attempt not only to summarize the materials scattered in the literature, but also give a discussion of the

<>
This work was supported in part by the U.S. NIH grants R21EB020861 and R01EB025133, National Natural Science Foundation of China U1805261 and National Key R&D Program of China 2017YFC0108802.

D. Liang and J. Cheng are with the Paul C. Lauterbur Research Center for Biomedical Imaging, D. Liang and Z. Ke are with the Research Center for Medical AI, Shenzhen Institutes of Advanced Technology, Chinese academy of Sciences, Shenzhen, Guangdong, China. J. Cheng and Z. Ke are also with Shenzhen College of Advanced Technology, University of Chinese Academy of Sciences, Shenzhen, China. (e-mail: {dong.liang, jing.cheng, zw.ke}@ siat.ac.cn).

L. Ying is with the Departments of Biomedical Engineering and Electrical Engineering, University at Buffalo, the State University of New York, Buffalo, NY 14260 USA (e-mail: leiying@buffalo.edu).

#D. Liang and J. Cheng contribute equally to this paper.

*indicates the corresponding author.

**Table 1.** Relationship between data-driven and model-driven approaches.

|  | Data-driven | Model-driven |
|---|---|---|
| Physical model | No | Based on |
| Architecture | Standard (e.g., U-net) | Dependent on a specific algorithm |
| Training algorithm | Back-propagation | Back-propagation |
| Training data | Relatively large | Relatively small |

**Table 2.** The notations used in the paper.

| | |
|---|---|
| $\mathbf{m}$: MR image, $\mathbf{m} \in \mathbb{C}^Y$ | $L$: the number of regularization functions |
| $\mathbf{A}$: encoding matrix, $\mathbf{A}: \mathbb{C}^Y \to \mathbb{C}^X$ | $\mathbf{K}$: convolution kernel |
| $\mathbf{f}$: undersampled k-space data, $\mathbf{f} \in \mathbb{C}^X$ | $N$: number of iterations |
| $\mathbf{D}$: transform matrix | $N_s$: number of training samples |
| $G$: regularization function | $n$: $n$-th iteration |
| $g$: nonlinear operator | $\mathcal{C}$: CNN block |
| $H$: activation function | $\widehat{\Theta}$: parameter set of the network |

relationship among these methods. It by no means includes a complete list of references of all contributions, as the field is fast growing, but the methods introduced here should serve as good examples to understand the field.

The paper is organized as follows. Section II gives a brief introduction of deep learning and MRI reconstruction basics. The classical deep learning network architectures and the general formulation of CS-based MRI reconstruction (from undersampled measured data to image) are provided. Section III introduces model-driven deep networks, with a focus on how each of the popular iterative, CS-based reconstruction algorithms is unrolled to a network with a unique architecture. In particular, the regularization parameters and functions to be learned via training are highlighted. Section IV introduces the data-driven methods. Networks with additional domain knowledge are elaborated. Section V reveals the connections between the two types of networks and introduces the third type – integrated approaches. Section VI discusses some signal processing issues and possible future directions. Section VII provides concluding remarks.

## II. BASICS OF DEEP LEARNING AND MRI RECONSTRUCTION

### A. Deep learning

Deep learning is a class of machine learning algorithms that exploits many cascaded layers of nonlinear information processing units to learn the complex relationships among data. By going deeper (i.e. more layers), the network improves its capability of learning features from higher levels of the hierarchy formed by the composition of lower level features. Such capability of learning features at multiple levels of abstraction allows the deep network to learn the complex functions that map the input to the output directly from data, without depending on human-crafted features.

The core of deep learning is deep neural network, which actually belongs to artificial neural network (ANN). Although ANN was invented in 1950s [27] and there were several ANN craze earlier, the latest one, known as deep learning, fully exhibits its power in various fields. There are three key factors that contribute to the recent success of deep learning: a) the invention of the optimization algorithms (e.g., layer-wise pre-training, mini-batch stochastic gradient descent (SGD), rectified linear units (ReLU), batch normalization, shortcut et.al.) enables training of high-dimensional, multivariate models; b) the availability of large datasets (Big Data) overcomes the overfitting issue, thanks to cloud storage; c) the ever-growing computational power of hardware (e.g., GPU and parallel computing) allows training performed in finite time, thanks to Moore's Law. In addition, the availability of open source software library (e.g., TensorFlow, PyTorch, Caffe, MatCovNet) makes the development of deep learning methods more efficient.

Most existing deep learning algorithms use ANN with supervised learning. In supervised learning, the weights and biases are learned from training data by minimizing a loss function (e.g., root mean square error, cross entropy, etc.). During training, a back-propagation algorithm is used to calculate the gradients of a loss function with respect to each weight/bias. Those gradients are then used in optimization algorithms (e.g., SGD) to update the weights/biases in a direction opposite the gradient accordingly. Updating the weights/biases multiple times on different training samples will eventually result in a properly trained neural network.

Classic types of deep neural networks include multilayer perceptron (MLP), convolutional neural networks (CNN), recurrent neural networks (RNN), and generative adversarial networks (GAN).

#### 1) MLP

MLP is one of the simplest neural networks. It consists of an input layer, an output layer, and at least one hidden layer. In a network with fully connected nodes between adjacent layers, each node, known as a neuron, uses a nonlinear activation function to operate on the sum of the biases and weighted outputs of connected nodes from the previous layer, thereby allowing representation of complex functions.

#### 2) CNN

The CNN has been mostly successful in image processing due to capability to learn position and scale invariant structures in the data, which is important when working with images. This technique is seeing great success because it scales with data and model size, and can be trained with back-propagation, but with a fraction of the computational complexity of MLP.

An $N$-layer CNN $\mathcal{C}_N = \mathcal{C}(m_0; \widehat{\Theta})$ with input $m_0$ and output $\mathcal{C}_N$ can be described as follows

$$\begin{cases} \mathcal{C}_0 = m_0 \\ \mathcal{C}_n = H_n(\Omega_n * \mathcal{C}_{n-1} + b_n) & n \in \{1,2,\dots,N-1\}, \\ \mathcal{C}_N = \Omega_N * \mathcal{C}_{N-1} + b_N \end{cases} \quad (1)$$

where $\Omega_n$ denotes the convolution operator of size $k_{n-1} \times w_n \times h_n \times k_n$ and $b_n$ is the $k_n$ dimensional bias with its element associated with a filter. The CNN output is the output of the final layer. Here, $k_{n-1}$ is the number of the image features extracted at the layer *n*-1, $w_n \times h_n$ means the filter size and $k_n$ is filter number at layer *n*, and $H_n$ means the nonlinear mapping operator. Eq. (1) could be regarded as the forward pass of the CNN training, where the convolution operator $\Omega_n$ is used to extract the features and $H_n$ is the nonlinear activation function. For example, the widely used activation function ReLU is given as:

$$H(x) = \begin{cases} x, & x \geq 0 \\ 0, & otherwise \end{cases} \quad (2)$$

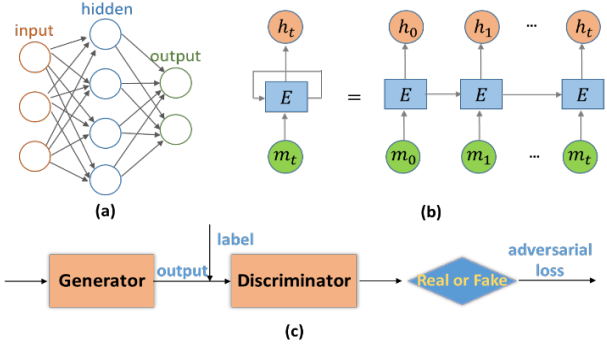

Figure 1. (a) Architecture of MLP with three layers. (b) Architecture of RNN. (c) Architecture of GAN.

Several advanced architectures have been integrated into CNN, such as U-net and ResNet to benefit information preservation and network optimization.

*3) RNN*

RNN is a class of neural networks that makes use of sequential information to process sequences of inputs. They maintain an internal state of the network acting as a "memory", which allows RNNs to naturally lend themselves to the processing of sequential data.

In Fig. 1(b), a chunk of neural network, $E$, takes some input $m_t$ and outputs a value $h_t$. Each $E$ loop allows information to be passed from one step of the network to the next. A recurrent neural network can be thought of as multiple copies of the same network, each passing a message to a successor.

*4) GAN*

GAN has gained popularity due to its ability to infer photo-realistic natural images. In GAN, there are two sub-networks, a generator $\mathcal{G}$ and a discriminator $\mathcal{D}$. The generator can generate high perceptual quality images according to the discriminator, which is a very good classifier to separate realistic and generated images. The loss function used for the generator is a combination of a content term and an adversarial term:

$$loss_G = loss_{cont} + \lambda loss_{adv}, \quad (3)$$

where λ is a hyper-parameter. The first term is a traditional loss function using mean squared error (MSE) between pixels,

$$loss_{cont} = \frac{1}{N_s}\sum_{j=1}^{N_s}\left\|\mathcal{G}(\mathbf{m}_j;\theta) - \mathbf{m}_j^{ref}\right\|_2^2, \quad (4)$$

where $N_s$ is the number of training samples, $\mathbf{m}_j^{ref}$ the ground truth image, $\mathcal{G}(\mathbf{m}_j;\theta)$ is the output of the generator of training set *j*. The term $loss_{adv}$ is defined based on the probabilities of the discriminator $\mathcal{D}$ over all training datasets as

$$loss_{adv} = \sum_{j=1}^{N_s}\log\left(1 - \mathcal{D}\left(\mathcal{G}(\mathbf{m}_j;\theta)\right)\right), \quad (5)$$

where $\mathcal{D}\left(\mathcal{G}(\mathbf{m}_j;\theta)\right)$ is the probability that the reconstructed image is a ground truth image.

*B. MRI reconstruction*

In MRI, the spatial information of the subject such as spin density and relaxation parameters is encoded in the measured data in a variety of ways [28, 29]. Typically, this encoding model is a mathematical description of how the MR signal is formed and how the spatial information is encoded into this signal. A linear encoding model is typically used after some approximation and can be written as

$$\mathbf{Am} = \mathbf{f}, \quad (6)$$

where **A** is the encoding matrix (e.g., Fourier and/or sensitivity encoding), **m** is the image and **f** is the measured data.

During data acquisition, the measured data **f** and the encoding model captured in the matrix **A** are known. The main task of the MR reconstruction is to recover the desired image from measured **f**. In some applications, MR reconstruction, is an under-determined inverse problem due to undersampling.

There are several methods to find solution **m** to Eq. (6), either by finding the least squares solution directly or through an iterative reconstruction scheme. For example, with certain assumptions, the encoding can be formulated as Fourier encoding. The acquired data is then called k-space data. If the Nyquist sampling criterion is satisfied, the image can be reconstructed by applying an inverse Fourier transform. In the case of Cartesian sampling, fast Fourier Transform (FFT) can be used for efficient direct reconstruction. Parallel imaging entails more complicated encoding matrices but also can be reconstructed by direct matrix inversion. For non-Cartesian or sub-Nyquist sampling, iterative reconstruction has to be used. In such scenarios, domain knowledge is often incorporated in the reconstruction model to facilitate the reconstruction, such as the spatio-temporal correlation in dynamic imaging and the established model of MR quantitative parameters.

CS is one of the revolutionary approaches to reconstruction from sub-Nyquist sampled data. The method exploits some prior models such as sparsity and low-rankness and solves the underlying constrained optimization problem. Details can be found in other papers in this special issue. In general, the imaging model from sub-Nyquist data can be written as

$$\hat{\mathbf{m}} = \underset{\mathbf{m}}{\mathrm{argmin}}\|\mathbf{Am} - \mathbf{f}\|_2^2 + G(\mathbf{m}), \quad (7)$$

where $\mathbf{A}: \mathbb{C}^Y \to \mathbb{C}^X$ is the encoding matrix. For example, in Fourier encoding model, and $\mathbf{A}=\mathbf{F_u}$ for single channel acquisition, $\mathbf{A}= \mathbf{F_u}\,\mathbf{S}$ for multi-channel acquisition with the coil sensitivity **S** and undersampled Fourier transform $\mathbf{F_u}$, where $\mathbf{f} \in \mathbb{C}^X$ is the acquired k-space data, $\mathbf{m} \in \mathbb{C}^Y$ is the image to be reconstructed, and $G(\mathbf{m})$ denotes a combination of sparse, low-rank, and/or other types of regularization functions.

In CS, the sparsity prior is usually enforced by fixed sparsifying transforms or data-driven but linear dictionaries. In contrast, deep learning goes beyond CS by extending the key ingredients of CS: adaptive sparsity and nonlinearity of the representation. It also addresses the issue of high computational complexity in CS reconstruction.

Table 3. Summary of some model-driven and data-driven methods.

| | method | model | algorithm | network | learning issues | remarks |
|---|---|---|---|---|---|---|
| model-driven | ADMM-net [24] | $\|\mathbf{Am}-\mathbf{f}\|_2^2 + \sum_{l=1}^{L} \lambda_l g(\mathbf{D}_l \mathbf{m})$ | ADMM | CNN | parameters, transform $D_l$, nonlinear operator $g$. | Single-coil imaging |
| | Variational-net [25] | $\|\mathbf{Am}-\mathbf{f}\|_2^2 + \sum_{l=1}^{L} \langle g_l(\mathbf{D}_l \mathbf{m}), \mathbf{1}\rangle$ | Gradient Decent | CNN | parameters, transform $D_l$, nonlinear operator $g_l$. | Multi-coil imaging |
| | ISTA-net [26] | $\|\mathbf{Am}-\mathbf{f}\|_2^2 + \lambda\|\mathbf{Dm}\|_1$ | ISTA | CNN | parameters, transform $D$. | Single-coil imaging |
| data-driven | AUTOMAP [8] | -- | -- | MLP, CNN | undersampled k-space → reference image | All reconstruction task |
| | GANCS [10] | -- | -- | GAN | aliased image → reference image | -- |
| | RAKI [13] | -- | -- | CNN | undersampled k-space → full k-space | Parallel imaging, database-free |
| | QSMnet [32] | -- | -- | U-net | local field maps → QSM maps | QSM |
| | DRONE [33] | -- | -- | MLP | MRF data → quantitative parameter values | MR fingerprinting |

## III. MODEL-DRIVEN DEEP LEARNING FOR FAST MR

### A. Definition

Although deep learning has gained a lot of success, the selection of network topology is still an engineering problem instead of a scientific research. In most existing deep learning approaches, there is a lack of theoretical explanation on the relationship between the network topology and performance. In addition, the generalizability of most networks is not understood. These are the common limitations of deep learning approaches.

Extending upon model-based CS methods, model-driven deep learning reconstruction methods provide a possibility to understand the relationship between network topology and performance. There are three ingredients in a model-driven deep learning method: model, optimization algorithm and deep network. A mathematical model between the measured data and the reconstructed image is first constructed based on the MR physics and other prior knowledge. It consists of free parameters and functions with a corresponding solution space. An optimization algorithm is then designed to reconstruct the image from the measured data based on the model and the convergence of the algorithm is established. The algorithm is finally unrolled to a deep network in which all free parameters and functions in the model are learnable using training data. In this way, the topology of the deep network is determined by the algorithm, and the deep network can be trained through back-propagation on training data.

### B. Examples

The major difference among the model-driven methods lies in the architectures of network, which are derived from different optimization algorithms. In this section, we will review several key model-driven deep learning methods for fast MR imaging including ADMM-net, Variational-net and ISTA-net [23-26]. A summary of these methods is given in part of Table 3.

#### 1) ADMM-net

ADMM-net was designed by unrolling the Alternating Direction Method of Multipliers (ADMM) algorithm. It was first introduced in NIPS 2016 by Yang *et al*. [23] to address the drawbacks of CS-MRI reconstruction. The original network, denoted as basic-ADMM-CSNet here, learns the regularization parameters in the ADMM algorithm. The network was then generalized in their follow-up work [24], denoted as Generic-ADMM-CSNet. The improved network learns the image transformations and nonlinear operators used for the regularization function.

In the context of compressed sensing MRI reconstruction using Eq. (7), the regularization term can be written as

$$G(\mathbf{m}) = \sum_{l=1}^{L} \lambda_l\, g(\mathbf{D}_l \mathbf{m}), \qquad (8)$$

where $\mathbf{D}_l$ denotes a transformation matrix (e.g., discrete wavelet transform for a sparse representation), $g(.)$ is a nonlinear operator (e.g., $l_q$-regularizer (q ∈ [0,1]) to promote sparsity), $\lambda_l$ is the regularization parameter, and the number of regularization functions is $L$.

In basic-ADMM-CSNet, a set of independent auxiliary variables $\mathbf{z} = \{\mathbf{z}_1, \mathbf{z}_2, \cdots, \mathbf{z}_L\}$ is introduced in the transform domain that $\mathbf{z}_l = \mathbf{D}_l \mathbf{m}$. The transform $\mathbf{D}_l$ aims to sparsify the image, such that $\mathbf{z}_l$ is sparse, which makes the model a synthetic model. The ADMM iterations can thus be written as

$$\begin{cases} \underset{\mathbf{m}}{\mathrm{argmin}}\, \frac{1}{2}\|\mathbf{Am}-\mathbf{f}\|_2^2 + \sum_{l=1}^{L} \frac{\rho_l}{2}\|\mathbf{D}_l\mathbf{m}+\boldsymbol{\beta}_l-\mathbf{z}_l\|_2^2 \\ \underset{\mathbf{z}}{\mathrm{argmin}}\, \sum_{l=1}^{L}\left[\lambda_l g(\mathbf{z}_l) + \frac{\rho_l}{2}\|\mathbf{D}_l\mathbf{m}+\boldsymbol{\beta}_l-\mathbf{z}_l\|_2^2\right] \\ \underset{\boldsymbol{\beta}}{\mathrm{argmin}}\, \sum_{l=1}^{L} \langle \boldsymbol{\beta}_l, \mathbf{D}_l\mathbf{m}-\mathbf{z}_l\rangle \end{cases} \qquad (9)$$

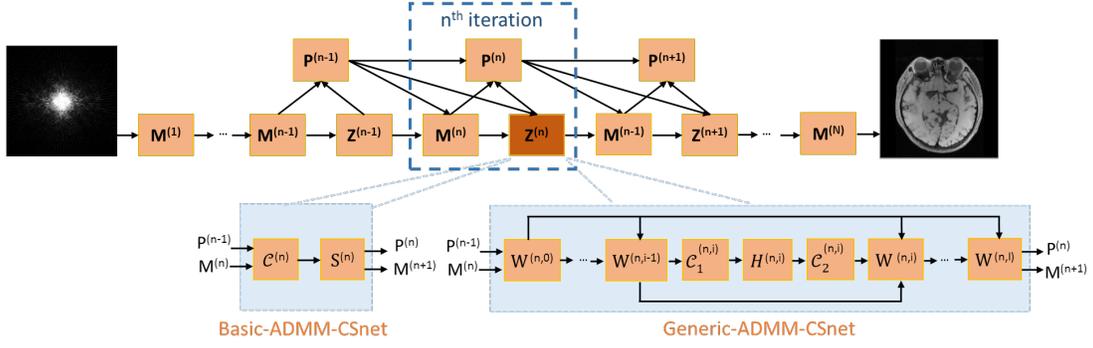

Figure 2. The data flow graph for the ADMM-net. $\mathcal{C}$: convolution; $H, S$: nonlinear operator; W: addition.

The solution is

$$\begin{cases} \mathbf{M}^{(n+1)}: \mathbf{m}^{(n+1)} = \dfrac{[\mathbf{A}^T\mathbf{f} + \sum_{l=1}^{L} \rho_l \mathbf{D}_l^T(\mathbf{z}_l^{(n)} - \boldsymbol{\beta}_l^{(n)})]}{[\mathbf{A}^T\mathbf{A} + \sum_{l=1}^{L} \rho_l \mathbf{D}_l^T\mathbf{D}_l]} \\ \mathbf{Z}^{(n+1)}: \mathbf{z}_l^{(n+1)} = S(\mathbf{D}_l\mathbf{m}^{(n+1)} + \boldsymbol{\beta}_l^{(n)}; \lambda_l/\rho_l) \\ \mathbf{P}^{(n+1)}: \boldsymbol{\beta}_l^{(n+1)} = \boldsymbol{\beta}_l^{(n)} + \eta_l^{(n+1)}(\mathbf{D}_l\mathbf{m}^{(n+1)} - \mathbf{z}_l^{(n+1)}) \end{cases}, \quad (10)$$

where $S(\cdot)$ is a nonlinear proximal operator of $g(.)$ with parameters $\lambda_l/\rho_l$.

In Generic-ADMM-CSNet, the independent auxiliary variables $\mathbf{z} = \{\mathbf{z}_1, \mathbf{z}_2, \cdots, \mathbf{z}_L\}$ is introduced in the spatial domain instead. Therefore, $\mathbf{D}_l\mathbf{z}$ is sparse, which makes the model an analysis model. The iterations of ADMM algorithm become

$$\begin{cases} \underset{\mathbf{m}}{\operatorname{argmin}} \dfrac{1}{2} \|\mathbf{A}\mathbf{m} - \mathbf{f}\|_2^2 + \dfrac{\rho}{2} \|\mathbf{m} + \boldsymbol{\beta} - \mathbf{z}\|_2^2 \\ \underset{\mathbf{z}}{\operatorname{argmin}} \sum_{l=1}^{L} \lambda_l g(\mathbf{D}_l\mathbf{z}) + \dfrac{\rho}{2} \|\mathbf{m} + \boldsymbol{\beta} - \mathbf{z}\|_2^2 \\ \underset{\boldsymbol{\beta}}{\operatorname{argmin}} \sum_{l=1}^{L} \langle \boldsymbol{\beta}, \mathbf{m} - \mathbf{z} \rangle \end{cases}, \quad (11)$$

yielding the solution:

$$\begin{cases} \mathbf{M}^{(n+1)}: \mathbf{m}^{(n+1)} = \dfrac{[\mathbf{A}^T\mathbf{f} + \rho(\mathbf{z}^{(n)} - \boldsymbol{\beta}^{(n)})]}{(\mathbf{A}^T\mathbf{A} + \rho\mathbf{I})} \\ \mathbf{Z}^{(n+1)}: \mathbf{z}^{(n+1,i+1)} = \mu_1 \mathbf{z}^{(n,i)} + \mu_2(\mathbf{m}^{(n+1)} + \boldsymbol{\beta}^{(n)}) \\ \quad - \sum_{l=1}^{L} \tilde{\lambda}_l^{(n+1)}(\mathbf{D}_l^{(n+1)})^T H_l^{(n+1)}(\mathbf{D}_l^{(n+1)} \mathbf{z}^{(n,i)}) \\ \mathbf{P}^{(n+1)}: \boldsymbol{\beta}^{(n+1)} = \boldsymbol{\beta}^{(n)} + \tilde{\eta}^{(n+1)}(\mathbf{m}^{(n+1)} - \mathbf{z}^{(n+1)}) \end{cases}, (12)$$

where $H(\cdot)$ refers to a nonlinear operation corresponding to the gradient of the regularizer $g(.)$, $i$ denotes the $i$-th iteration in each $\mathbf{Z}^{(n+1)}$.

In both Eqs. (10) and (12), the image $\mathbf{m}$, auxiliary variables $\mathbf{z}$, and multiplier $\boldsymbol{\beta}$ are updated iteratively. Although the computation performed in each iteration of both methods may be slightly different due to different representations of auxiliary variables, the data flow graph of the above two ADMM-nets is the same, which is illustrated in Fig. 2. As shown in Fig. 2, the $n$-th iteration in the data flow graph corresponds to $n$-th iteration of the ADMM algorithm. Specifically, in each iteration of the graph, there are three types of nodes corresponding to three modules in ADMM: reconstruction module ($\mathbf{M}$), denoising module ($\mathbf{Z}$), and multiplier update module ($\mathbf{P}$). In basic-ADMM-CSnet, all parameters ($\rho_l, \lambda_l, \eta_l$) in the original ADMM algorithm are learnable, while Generic-ADMM-CSNet also learns the image transformation $\mathbf{D}_l$, which is implemented linearly by convolving with kernels, and the nonlinear operator $g(.)$, which is realized by a piecewise linear function determined by a set of control points.

*2) Variational-net*

The aforementioned ADMM-nets are designed for the single coil MR reconstruction, which limits their application in clinical settings where most scans use phased-array coils. The variational network (Variational-net) [25] was developed for multi-coil MR image reconstruction. It combines the mathematical structure of variational models for CS reconstruction with deep learning to learn the regularization parameters, image transformations, and nonlinear operators.

The regularization term $G(\mathbf{m})$ in Eq. (7) is defined by the Field of Experts model in a variational model:

$$G(\mathbf{m}) = \sum_{l=1}^{L} \langle g_l(\mathbf{D}_l\mathbf{m}), \mathbf{1} \rangle, \quad (13)$$

where $\mathbf{1}$ denotes a vector of all ones. Each linear transform $\mathbf{D}_l$ is realized by convolution with filter kernels $\mathbf{K}_l$. The regularization parameters are implicitly contained in $g(.)$. With the Field of Experts model, the solution to Eq. (7) is iteratively updated in the gradient decent direction as

$$\mathbf{m}^{(n+1)} = \mathbf{m}^{(n)} - \sum_{l=1}^{L} (\mathbf{D}_l^{(n)})^T H_l^{(n)}(\mathbf{D}_l^{(n)}\mathbf{m}^{(n)}) \\ - \tilde{\lambda}^{(n)} \mathbf{A}^T(\mathbf{A}\mathbf{m}^{(n)} - \mathbf{f}), \quad (14)$$

where $H_l^n$ is the activation function corresponding to the first order derivative of the nonlinear operator $g_l$. The step size of gradient decent is implicitly contained in the activation functions and the data fidelity weights $\tilde{\lambda}$. Here, the encoding matrix A consists of sub-Nyquist Fourier encoding and coil sensitivity.

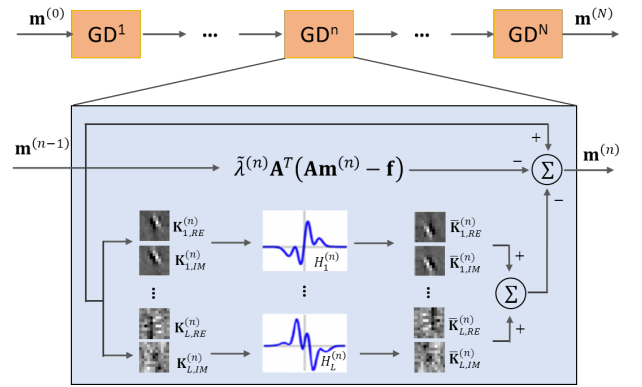

Figure 3. Structure of Variational-net.

Variational-net is obtained by unfolding the iterations of Eq. (14), which is depicted in Fig. 3. As MR data is complex, the convolution includes a real part and an imaginary part. The transpose operation $(\mathbf{D}_l^n)^T$ in Eq. (14) can be implemented as a convolution with the filter kernels of $\mathbf{D}_l$ but rotated by $180^0$. The parameters learned in Variational-net are the filter kernels $\mathbf{K}_l^n$, activation functions $H_l^n$ and data weights $\tilde{\lambda}^{(n)}$.

It is worth noting that the sensitivity maps need to be pre-calculated and fed into Variational-net.

*3) ISTA-net*

ISTA-net [26] aims to solve the general CS reconstruction problem with an unrolled version of the iterative shrinkage-thresholding algorithm (ISTA). It uses deep learning to learn the image transformation and parameters involved in original algorithm of ISTA.

ISTA solves the optimization problem in Eq. (7) as
$$\mathbf{r}^{(n+1)} = \mathbf{m}^{(n)} - \rho \mathbf{A}^T(\mathbf{A}\mathbf{m}^{(n)} - \mathbf{f}) \quad (15)$$
$$\mathbf{m}^{(n+1)} = \underset{\mathbf{m}}{\operatorname{argmin}} \frac{1}{2} \left\| \mathbf{m} - \mathbf{r}^{(n+1)} \right\|_2^2 + G(\mathbf{m}), \quad (16)$$
where $G(\mathbf{m}) = \lambda \|\mathbf{Dm}\|_1$ is a regularization function and $\rho$ is the step size. It is difficult to obtain $\mathbf{m}^{(n+1)}$ if the transformation $\mathbf{D}$ is non-orthogonal or even nonlinear.

ISTA-net is an unrolled version of the traditional ISTA of Eq. (15) and Eq. (16), but overcomes the abovementioned drawback. With a general form of image transformation $\mathbf{D}(\mathbf{m})$, the $\mathbf{m}^{(n+1)}$ module in Eq. (16) becomes
$$\mathbf{m}^{(n+1)} = \underset{\mathbf{m}}{\operatorname{argmin}} \frac{1}{2} \left\| \mathbf{D}(\mathbf{m}) - \mathbf{D}(\mathbf{r}^{(n+1)}) \right\|_2^2 + \theta \|\mathbf{D}(\mathbf{m})\|_1, \quad (17)$$
where $\theta$ is the merged parameter related to $\lambda$ and $\mathbf{D}(\cdot)$. The image can be obtained as
$$\mathbf{m} = \widetilde{\mathbf{D}}\left(soft(\mathbf{D}(\mathbf{r}^{(n+1)}), \theta)\right), \quad (18)$$
where $\widetilde{\mathbf{D}}(\cdot)$ is the left inverse of $\mathbf{D}(\cdot)$.

The architecture of the ISTA-net is illustrated in Fig. 4. Each iteration of ISTA-net consists of two modules: $\mathbf{r}^{(n+1)}$ module and $\mathbf{m}^{(n+1)}$ module. The $\mathbf{r}^{(n+1)}$ module is the same as Eq. (15) except that the step size $\rho$ is learnable in each iteration. In the $\mathbf{m}^{(n+1)}$ module, $\mathbf{D}(\cdot)$ is modeled as two convolutional operators separated by a ReLU operator as depicted in Fig. 4, $\widetilde{\mathbf{D}}(\cdot)$ exhibits a structure symmetric to that of $\mathbf{D}(\cdot)$.

ISTA-net is designed for the general CS reconstruction problem, not MR reconstruction only. Different from ADMM-net and Variational-net, ISTA-net adopts an $l_1$-regularizer for the sparse prior, which is restricted by ISTA.

## IV. DATA-DRIVEN DEEP LEARNING FOR FAST MR

Data-driven deep learning methods originally employ the standard neural networks to directly learn the mapping between input (undersampled k-space data or aliased images) and output (clean images) without utilization of the domain knowledge. In contrary to model-driven approaches, the networks of data-driven approaches cannot be understood as the unrolling of a specific iterative algorithm. Usually, it requires a large size of training sets, high computational power and engineering experience to design and train the network, which may limit the practical applications of data-driven methods. An alternative way to lower the prerequisites and improve the performance is to add some prior knowledge into the network.

In this section, we will review some basic data-driven deep learning methods for fast MR imaging (some are summarized in Table 3) and introduce some additional domain knowledge from MRI to be incorporated into data-driven methods [6-16, 30-33].

### A. Basic data-driven networks for MR reconstruction

The idea of using CNN for MR reconstruction was first proposed by Wang *et al* [6]. It directly uses the structure of CNN (1) to learn the network-based mapping between the aliased images and the clean images. In most existing works, the aliased image which is the inverse Fourier transform from undersampled k-space data is used as the network input and the desired image from the fully sampled k-space data as the network output. The mapping is learned through standard networks such as MLP, U-net and ResNet.

AUTOMAP [8] uses a fully-connected network as the first layer to perform the Fourier transform in MR reconstruction as Fourier transform is a global transform. The k-space data is directly taken as the network input and the reconstructed image can be obtained as network output. RAKI [13] uses the 3-layer CNN to learn the k-space interpolation for parallel imaging.

Due to its outstanding performance in image-to-image translation, GAN has been exploited in MRI reconstruction [9, 10] to correct the aliasing artifacts in the zero-filled reconstruction from undersampled MR data.

The deep networks are also used in the estimation of quantitative tissue parameters from recorded complex-valued data. Such as in MR quantitative parameter mapping [30, 31], quantitative susceptibility mapping [32], and magnetic resonance fingerprinting [33]. QSMnet [32] constructs a 3D CNN with the architecture of U-Net to generate high quality susceptibility source maps from single orientation data. And DRONE [33] adopts a 4-layer MLP to extract tissue properties and predict quantitative parameter values (T1 and T2) from 2D MRF data.

### B. Domain Knowledge from MRI

Although direct application of standard neural networks to MRI reconstruction has been the first attempt, further development with additional domain knowledge can improve the performance of standard networks.

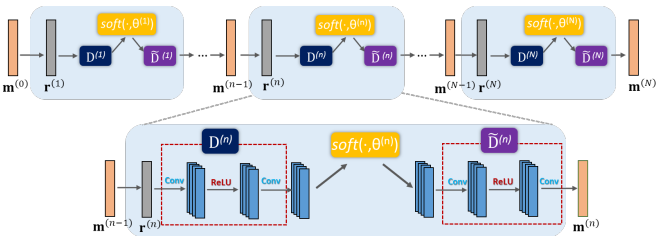

Figure 4. Architecture of the ISTA-net.

*1) Fourier transform*

Most existing data-driven approaches use the aliased image that is the zero-filled reconstruction of the undersampled k-space data as the input of the network. By doing that, the methods take advantage of the knowledge unique to MRI data. AUTOMAP [8] differs from most approaches in that it takes the k-space data directly as the network input and thus the network also needs to represent the Fourier transform. Although AUTOMAP provides a good direction for the true end-to-end networks, it requires huge memory to store the fully-connected layers. In addition, since analytical Fourier transformation can be performed efficiently through FFT, it might not be worthwhile to learn the transform using complex networks.

*2) Regularization term*

In [6], Wang *et al* proposed two options for applying the network-based approach with the CS-based one. One is to use the image reconstructed from the learned network as the initialization for compressed sensing method in Eq. (7). The other is to use the image generated by the network as a reference image in an additional regularization term. The formulation of the latter one can be described as

$$\hat{\mathbf{m}} = \arg\min_{\mathbf{m}} \left\{ \|\mathbf{Am} - \mathbf{f}\|_2^2 + G(\mathbf{m}) + \alpha \|\mathcal{C}(\mathbf{A}^T\mathbf{f}; \Theta) - \mathbf{m}\|_2^2 \right\} \quad (19)$$

where $\alpha$ is the regularization parameter. The difference between the latter option and model-driven methods is that the solution of (19) is accomplished by a neural network in the model-driven methods, whereas in [6], the solution is obtained by the conventional CS algorithms without any unrolling which is typical in the model-driven methods.

*3) Data consistency*

In traditional image reconstruction algorithms such as CS, data consistency in k-space is one of the most critical constraints. Such cross-domain prior information can also be very useful when incorporated in the data-driven deep learning methods. Since most learning methods use images as the input and output of the network, the images need to be transformed to the k-space to enforce data consistency. For example, Schemper *et al* [17] convert the intermediate reconstruction from the CNN to k-space to enforce data consistency. KIKI-net [18] alternates between the image domain (I-CNN) and k-space (K-CNN) iteratively where the data consistency constraint is enforced in an interleaving manner, as shown in Fig. 5. Each K-CNN is trained to minimize the loss defined as the difference between the reconstructed and fully sampled k-space data, thereby taking full advantage of all the k-space measurements. KIKI-Net addresses the issues that some structures are already lost in the network input if the CNN is only trained in the image domain with the aliased image as the input. It is effective in restoring detailed tissue structures in images as well as in reducing aliasing artifacts.

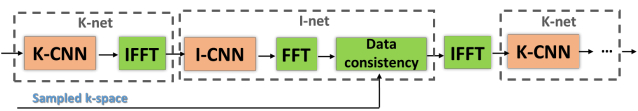

Figure 5. Architecture of KIKI-net.

*4) Spatio-temporal correlations*

In dynamic MRI, the temporal correlation has shown to improve the reconstruction quality of compressed sensing based methods. In order to take advantage of such domain knowledge, a residual U-net structure with 3D convolution has been used to suppress artifacts [15]. An alternative approach is employing a convolutional RNN to jointly exploit the dependencies of the temporal sequences and iterations [20]. Another approach exploiting the spatio-temporal correlations is to use data sharing where the k-space data is shared among neighboring frames along the temporal direction [17].

*5) Quantitative parameters*

Estimation of quantitative tissue parameters from recorded complex-valued data is another exciting field in deep learning MR imaging. The networks can be designed by incorporating the physical model of the quantitative parameters to be mapped [31].

## V. INTEGRATED DEEP LEARNING FOR FAST MR

### A. Connections between two approaches

The pure data-driven network initially proposed in [6] has the advantage of simplicity as the network treats image reconstruction as a "black box" of input to output without the need for any domain knowledge of MRI. However, such network has high demand on the quality and quantity of training samples. That is, the network is prone to the issues of underfitting and overfitting. After these limitations are identified, several methods were developed to incorporate domain knowledge in the data-driven networks, such as the examples in Section IV. As more and more domain knowledge is included in the network, the reconstruction quality is expected to improve.

On the other hand, the model-driven network roots from a basic compressed sensing reconstruction algorithm, where all regularization functions are pre-determined, and the regularization parameters are manually tuned. Deep learning was initially introduced to unroll the iterations of the reconstruction algorithms, allowing the regularization parameters to be learned [23]. A number of networks were further developed to relax the fixed constraints using learnable operators and functions, such as the examples in Section III. As the constraints in the reconstruction model are gradually relaxed and more functions in the model are learned, the reconstruction quality is expected to improve.

The data-driven approach usually requires more training data than the model-driven approach, although more development is going on to reduce the required training data for all methods. Based on the above observations, as the constraints are learned in the model-driven network and domain knowledge is incorporated in the data-driven network, the model-driven network and the data-driven network may eventually "meet" at a point where a specific deep learning reconstruction method can be explained from either perspective. For example, there are a few deep learning reconstruction methods that have the ingredients of both model-driven and data-driven. We call them integrated approaches.

## B. Integrated approaches for MR reconstruction

A method belonging to the integrated approach has two features: it is an unrolling version of an optimization algorithm (e.g. alternating minimization), which is typical of model-driven approaches, and at least one sub-problem of the algorithm is solved using the "black box", which is typical of data-driven approaches. Some examples including MoDL, DCCNN, KIKI-net, PD-net, CRNN-MRI, are given as follows.

### 1) MoDL

MoDL, model-based deep learning framework [19], uses the CNN unit as a denoiser. It formulates the regularization term in Eq. (7) as
$$G(\mathbf{m}) = \lambda \|\mathbf{m} - \mathcal{C}(\mathbf{m})\|^2,$$
where $\mathcal{C}(\mathbf{m})$ is the denoised image of $\mathbf{m}$ obtained by a CNN unit. With the alternating minimization algorithm, image can be reconstructed as followings:
$$\mathbf{m}^{(n+1)} = \underset{\mathbf{m}}{\operatorname{argmin}} \|\mathbf{Am} - \mathbf{f}\|_2^2 + \lambda \|\mathbf{m} - \mathbf{z}^{(n)}\|^2 \quad (20)$$
$$\mathbf{z}^{(n+1)} = \mathcal{C}(\mathbf{m}^{(n+1)}). \quad (21)$$
The data consistency (DC) sub-problem (20) is solved by the conjugate gradient (CG) algorithm to handle more complex forward models (e.g. multichannel MRI). The CG process needs not to be unrolled to networks as there are no trainable parameters in it. And DC is integrated as a layer into the network. The denoised image (21) is achieved through the CNN unit without any prior knowledge.

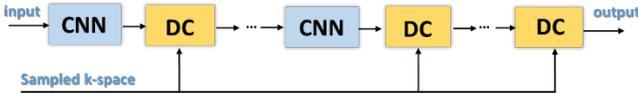

Figure 6. Network architecture of MoDL and DCCNN.

### 2) DCCNN

DCCNN, a deep cascade of convolutional neural networks for dynamic MR reconstruction [17], improves the CNN-based reconstruction by adding data consistency. It also iterates Eqs. (20) and (21) to reconstruct the images. Different from MoDL which enforces data consistency by integrating the CG algorithm with the network to suit multichannel acquisitions, DCCNN considers single channel acquisition such that Eq. (20) has a close-formed solution:
$$\mathbf{m} = (1 + \lambda \mathbf{A}^T \mathbf{A})^{-1}(\mathcal{C}(\mathbf{m}_0; \Theta) + \lambda \mathbf{A}^T \mathbf{f}) \quad (22)$$
The architecture of DCCNN shares similarities with the "black box" CNN unit, but can also be viewed as unrolling an alternating minimization algorithm with CNN-derived constraints. The data-driven with cross domain learning method KIKI-net [18] also shows this property.

### 3) PD-net

PD-net, the learned primal dual proposed by Adler and Öktem for tomographic reconstruction [21], has been applied in MRI reconstruction [22]. PD-net solves a minimization formulation as
$$\min F(\mathbf{Am}) + G(\mathbf{m}). \quad (23)$$
It is worth noting that Eq. (7) is a special case of Eq. (23) if $F(\mathbf{Am}) = \|\mathbf{Am} - \mathbf{f}\|_2^2$. Introducing an auxiliary dual variable $\mathbf{d}$, the optimization problem in Eq. (23) can be solved by the primal dual hybrid gradient algorithm as
$$\begin{cases} \mathbf{d}_{n+1} = prox_\alpha[F^*](\mathbf{d}_n + \alpha \mathbf{A}\bar{\mathbf{m}}_n) \\ \mathbf{m}_{n+1} = prox_\tau[G](\mathbf{m}_n - \tau \mathbf{A}^* \mathbf{d}_{n+1}) \\ \bar{\mathbf{m}}_{n+1} = \mathbf{m}_{n+1} + \theta(\mathbf{m}_{n+1} - \mathbf{m}_n) \end{cases} \quad (24)$$
where $\alpha$, $\tau$ and $\theta$ are the algorithm parameters, $prox$ denotes the proximal operator. In PD-net, the proximal operators are replaced with parameterized operators Γ and Λ where the parameters can be learned using training data, and the fixed variable combinations inside the operators Γ and Λ are also relaxed. Therefore, the PD-net can be formulated as
$$\begin{cases} \mathbf{d}_{n+1} = \Gamma(\mathbf{d}_n, \mathbf{Am}_n, \mathbf{f}) \\ \mathbf{m}_{n+1} = \Lambda(\mathbf{m}_n, \mathbf{A}^* \mathbf{d}_{n+1}) \end{cases} \quad (25)$$
The architecture of the PD-net is shown in Fig. 7. It is worth mentioning that the data fidelity term in the PD-net is no longer the $l_2$ norm of the estimated error at the sampled locations as in Eq. (7), but is learned implicitly from the training data instead, and the regularization term is also learned implicitly.

PD-net is an unrolling version of the primal dual algorithm (24). On the other hand, the dual and primal sub-problems in Eq. (25) are solved by two CNN blocks without model. Particularly, to ensure data fidelity, PD-net directly conveys the sampled k-space to the k-space-based CNN rather than using the model-based minimization (20).

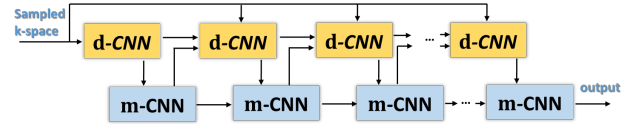

Figure 7. Architecture of PD-net.

The above examples can be viewed as the examples of the integration of model-driven and data-driven networks. Integrated approaches are the trend in deep learning MR reconstruction as they have the benefits of both - flexibility of models and powerful learning capability of deep networks. More techniques of "optimal" integration are to be developed in the future.

## VI. SOME SIGNAL PROCESSING ISSUES

### A. Theoretical analysis

Unlike constrained reconstruction with sparsity or low rank constraints, the theoretical guarantee for deep reconstruction is largely unexplored. The paper [12] provided a preliminary theoretical rationale for some existing deep learning architectures and components. It used the concept of convolutional framelet to explain deep CNN for inverse problems. It shows that deep learning is closely related to the annihilating filter-based approaches, where the lifted Hankel matrix usually results in a low-rank structure that can be decomposed using both non-local and local bases. Based on the framelet framework, in order for the deep network to satisfy the perfect reconstruction condition, the number of channels should increase exponentially with the layer, which is difficult to achieve in practice; when insufficient number of filter channels is used instead, the network is actually performing a low-rank based shrinkage. Therefore, the depth of the network depends on the rank of the signal and the length of the convolution filter. As more domain knowledge is included with the general CNN structure and the connection with unrolled iterations of optimization algorithms becomes clearer, the theoretical analysis of a deep network might become more comprehensive.

*B. Transfer learning*

Apparently, machine learning can only capture what it has seen. If there were a significant difference between the statistical features in training data and testing data, the trained network would fail for the testing case. For example, the training stage should be performed every time the readout changes between the training and testing sets. However, it is not always possible to include all cases in training. Knoll *et al* [34] studied the generalizability of the trained variational-net in the case of deviation between training and testing data in terms of image contrast, SNR, and image content. In particular, when the contrast and SNR of the training data are quite different from the testing data, deep learning-based reconstruction has substantial level of noise or yields slightly blurred images with some residual artifacts. In addition, a network that was trained from regular undersampled data is easily generalized to randomly undersampled data, although the other way would introduce residual artifacts. It is worth noting that the conclusion is based on the empirical results from a specific variation network in [25]. In depth studies in more general settings are still needed.

In the scenario of minor mismatch between training and testing data, transfer learning can be a solution. For example, Based on the similarity between projection MR and CT, Han *et al.* proposed to train the network using CT data and then adapt the network parameter for MR reconstruction with fine-tuning [11]. AUTOMAP proposed by Zhu *et al* learns the mapping between the Fourier measurement and reconstruction using training data simulated from natural images and then applies to MRI reconstruction with success [8]. In most cases of transfer learning, additional fine tuning is necessary to achieve performance nearly identical to the network trained directly in the testing domain with a large size of data.

*C. Relationship with other learning-based approaches*

Before deep learning is popular in MRI reconstruction, there have already been quite a few learning-based reconstruction methods, where the prior models for compressed sensing reconstruction are also learned from some training data (e.g., [35-38]). These methods differ from the deep learning-based methods in that 1) there is no network structure in learning; 2) the prior comes from a specific mathematical model; 3) few training data are needed. For example, in compressed sensing with dictionary learning [35, 36], a linear transformation is learned using training data (simulated data from theoretical model or low-resolution image). In compressed sensing with manifold learning [37, 38], a highly nonlinear prior of low-dimensional manifold is learned using training data. These methods can also be viewed as machine learning-based approaches (e.g., kernel principal component analysis and manifold learning are all well-known machine learning methods) and can also benefit from availability of large numbers of training samples.

*D. Other issues in deep learning approaches*

The forward model in MR imaging is more than Fourier model, therefore, deep leaning applications with other trajectories, the complex-valued nature of MR data and prospective validation should be taken into account. As MR data is complex-valued, most works separate the real and imaginary parts into two channels, whereas [14] trains the magnitude and phase networks separately. Even so, the complex-valued issue is not really addressed. Multi-channel acquisition is a standard technology in clinical scans. To handle the data from multi-coils, [19] and [25] convey the pre-calculated coil sensitivity into the networks, while [7] directly reconstructs the image from the multi-channel data through networks. Different from others, [13] learns the k-space interpolation from the auto calibration signal without a large amount of training data.

Cartesian sampling is commonly used in the research of deep learning MR reconstruction as the transform can be performed efficiently through FFT. However, non-Cartesian acquisition is less prone to motion artifacts and allows higher under-sampling factors compared to Cartesian acquisition. AUTOMAP [8] could reconstruct the image directly from the non-Cartesian samples, while [11] provides an alternative approach to reconstruct the image from radial trajectory with the help of domain adaptation from CT projection.

Prospective validation is very important to evaluate a deep learning method in a real application. The works in [11], [13] and [25] provide the prospective experiment results.

*E. Future Directions*

Radiomics, a hot field for years, aims to extract extensive quantitative features from medical images using data-mining algorithms and the subsequent analysis of these features for computer aided diagnosis (CAD). With the emergence of deep learning, deep neural networks have been successfully utilized in radiomics for disease diagnosis [39]. On the other hand, as discussed in the article, deep learning will significantly improve the reconstructed image quality, which will directly benefit CAD by making the features in training and testing images more prominent. Another way to improve the image quality is to optimize the data acquisition protocol. Deep learning can also contribute to this endeavor by learning the optimal acquisition protocol through training.

Since in radiomics, images are not for visual inspection, but are combined with other patient data to be analyzed by deep learning algorithms, it is conceivable that there exists a synergistic opportunity to integrate deep reconstruction and radiomics for optimal diagnostic performance [40]. Deep learning can be applied to the entire end-to-end workflow from data acquisition, image reconstruction, radiomics, to the final diagnosis report. Such end-to-end workflow has a significant potential to improve diagnostic, prognostic, and predictive accuracy.

Typically, the networks in MR reconstruction are trained in a fully supervised manner, self-supervised learning or even unsupervised learning could be exploited in the future.

## VII. Conclusion

In conclusion, deep learning has demonstrated potentials in MR image reconstruction. In only a few years, various networks have already been developed, taking advantage of the unique properties of MRI. The overview of existing methods will inspire new developments for improved performance. Deep reconstruction will fundamentally influence the field of medical imaging.

Acknowledgment

The authors would like to thank Dr. Qiegen Liu for his helpful discussions.

ACKNOWLEDGMENT

The authors would like to thank Dr. Qiegen Liu for his helpful discussions.